\documentclass[useAMS,usenatbib]{mn2e}
\usepackage{float}
\usepackage{graphicx}
\usepackage{amssymb}
\usepackage{amsmath}
\usepackage{datetime}
\usepackage[normalem]{ulem}
\usepackage{times}
\usepackage[export]{adjustbox}
\usepackage{soul}
\usepackage{color}
\usepackage{soul}

\definecolor{stan}{rgb}{0,0,1}

\definecolor{steve}{rgb}{0,1,0}

\newcommand{\beq}{\begin{equation}}
\newcommand{\eeq}{\end{equation}}
\newcommand{\beqa}{\begin{eqnarray}}
\newcommand{\eeqa}{\end{eqnarray}}

\newcommand{\vw}{V_w}
\newcommand{\vwsq}{V_w^2}

\title[Cluster outflow]{
A generalized semi-analytic model for the combined outflow from a cluster of stars with strong stellar winds}
\author[S.\ P.\ Owocki]{
Stanley P.\ Owocki$$\thanks{email: owocki@udel.edu}
\\
Bartol Research Insitute, 
Department of Physics \& Astronomy, 
University of Delaware, Newark, DE 19716 USA 
}

\begin{document}

\include{aas_macros}

\date{Accepted ?.  Received ?; in original form ?}

\maketitle

\label{firstpage}

\begin{abstract}
We derive semi-analytic solutions for the thermally driven cumulative outflow from a dense stellar cluster that consists of a large number of stars with strong stellar winds, under the  key assumption that their mass and energy deposition can be treated as a continuous source of thermalized wind material. Our approach provides explicit analytic forms for the flow critical radius and sound speed, which then allows for full solution by inward/outward integration from this critical radius. Application to previous models that assume either a power-law or exponential  mass deposition allows a direct comparison of their velocity solutions. We then obtain solutions for new models that have mass depositions with a gaussian or the empirically motivated forms derived by Plummer and Elson. Comparisons show that, when cast in terms of the radius scaled by the critical radius, the overall velocity laws are all quite similar, asymptotically approaching the energy-averaged stellar wind speed $V_w$, and anchored to passing through a common critical/sonic speed $v_c = V_w/2$. By deriving associated variations in outflow density and temperature, we obtain scaling forms for the expected X-ray emission, and show that radiative cooling generally represents only a small correction to the assumed cooling from adiabatic expansion.  An initial analysis of incomplete wind thermalization suggests this should likewise have modest effect on the overall outflow in dense clusters. The simplified semi-analytic method presented here can be readily applied to alternative mass depositions and so form a basis for interpreting results from detailed numerical simulations of outflows from young stellar clusters.
\end{abstract}



\section{Introduction}
\label{sec:sec1}
\label{sec:intro}

Young, massive stellar clusters generally contain a significant population of  luminous, massive stars with powerful, radiatively 
driven stellar winds.
In the few- million-year epoch before these massive stars undergo core collapse (leading either to supernovae explosions that disrupt the cluster, or a direct collapse to a black hole that terminates wind mass loss),
the collective deposition of
mass, momentum, and energy from their stellar winds into the intra-cluster medium  can lead  to a strong  cluster outflow, 
driven largely by the thermal energy generated from wind-wind collisions.
The resulting wind-blown bubble into the surrounding interstellar medium can then take a form analogous to the
single-star wind-bubbles analyzed by \citet{Weaver77}, wherein the supersonic outflow terminates in a strong shock.
For clusters this termination shock, along with the wind-wind shocks internal to the cluster, can heat gas to X-ray emitting temperatures 
of order a few keV, and also be sites of particle acceleration to much higher energies.
The latter provides a potential basis for interpreting the high-energy (up to a PeV!) gamma rays observed from some
young clusters
\citep{Aharonian19,Peron24,Pandey24}.
\citet{Morlino21} discuss how radio observations can constrain models of particle acceleration in termination shocks.
Early analyses of wind-blown bubbles include \citet{Pikelner68}, \citet{Avedisova72}, and \citet{Dyson72}.

A focus of much current research is on developing sophisticated 3D numerical simulations of the wind-wind collisions
that form the cluster outflow, along with the resulting wind blown bubble, 
\citep[see, e.g.,][and references therein]{Vieu24,Badmaev22},
with an aim to derive associated observational signatures in X-rays and gamma rays.
For the specific case of Cygnus OB, \citet{Vieu24} find the cluster  is too sparse to produce a global termination shock, as the stars with the most powerful winds are so widely separated that they each are enveloped in their own bubble and termination shock.
The analysis here is instead aimed at denser, more massive clusters for which mutual collisions effectively thermalize the wind outflows.

Building on models of multiple interacting supernovae by  \citet{Chevalier85},  \citet{Canto00}  assumed a sufficiently
large and dense population of mass-losing stars that the cumulative discrete contributions of mass and energy can be modeled in 
terms of a {\em continuous} mass deposition of fully thermalized stellar wind material.
Under the particularly idealized assumption that this mass loading follows a power-law in radius,
\citet{Rodriguez-Gonzalez07} derived fully {\em analytic} solutions for the radial variation of the cluster outflow speed,
along with associated variations in density and temperature.
Building on initial analyses  \citep{Silich04, Silich05},
\citet{Silich11} carried out numerical integrations for a model with an exponential mass deposition, obtaining a velocity solution that has
a more gradual acceleration; but like the analytic models, this also reached a terminal speed 
set by the energy-averaged terminal speed of the individual stellar winds, which
for massive stars can approach $V_w \approx 3000$\,km/s \citep{Puls08}.
The cluster cumulative mass loss is just set by the total lost by the individual stellar winds.

A longstanding issue regards the degree to which the turbulent interaction of individual stellar wind sources leads them to become  fully thermalized \citep{Rosen14},  and why the observed X-ray emission is weaker than expected from the \citet{Weaver77} and \citet{Chevalier85}
models if all the wind energy is thermalized \citep{Lopez11}.
\citet{Koo92} explored fast vs. slow wind scenarios and how the associated thermalization and density influence the ambient density and 
balance of  heating versus cooling.
\citet{Gupta18} modeled how cosmic rays produced in stellar wind bubbles could explain the weak X-ray emission observed in star clusters.
\citet{Lancaster21} used numerical simulations to model turbulent mixing of wind bubbles driven by stellar clusters.
\citet{Rosen22} demonstrated that turbulent mixing happens very quickly for wind blown bubbles during massive star formation, leading to rapid energy losses and smaller wind bubbles. 
\citet{Harper-Clark09} focused on modeling physical leakage of the hot shock-heated gas as an explanation for the weak emission of the Carina Nebula.
\citet{Rogers13} used simulations to explore the role of feedback from winds and supernovae on massive stellar clusters.
\citet{Gallegos-Garcia20} used  simulations of wind-blown bubbles to study how they drive Kolmogorov turbulence, following also how thermal energy is lost due to turbulent mixing.

The aim of the present paper is to generalize previous  treatments of cluster outflows, to solve for the radial variation of
speed, density and temperature under more general models for the mass deposition, and use these to derive
scaling laws for radiative losses and associated X-ray luminosity as a function of cumulative mass loss rate, stellar wind speed, and cluster size.

In terms of the basic conservation equations, section \ref{sec:sec21} derives the critical differential equation (\ref{eq:eomgg}) for the transonic outflow,  from which one can obtain explicit expressions for the critical radius and associated critical speed;
direct inward/outward integration from this critical radius yields full velocity solutions for any given mass deposition function.
We then use this to reproduce and compare  previous  solutions that assume either power-law (2.2) or exponential forms  (2.3) for the
mass deposition. Section \ref{sec:sec24} shows that radiative cooling is generally a small perturbative effect, allowing then a general application of the semi-analytic approach developed here for purely adiabatic outflows.

Section \ref{sec:sec3} then uses this adiabatic approach to derive new flow models for mass depositions that follow either a gaussian or empirical forms suggested by \citet{Plummer11} and \citet{Elson87}  for the density distribution of clusters, 
and  compares
these with the previous  exponential model.
Section \ref{sec:sec41} derives corresponding variations in temperature and density; section 4.2 uses these
 to obtain scalings for radiative losses;  and  section 4.3 derives the associated  X-ray emission.
Section \ref{sec:sec5} discusses extensions to account for  a finite pressure in the interstellar medium (\ref{sec:sec51}),  incomplete thermalization from wind-wind collisions (\ref{sec:sec52}), and radiative cooling in stellar wind termination shocks (\ref{sec:sec53}).
The final section \ref{sec:sec6} summarizes our results and outlines directions for future work.

\section{Analysis}
\label{sec:sec2}
\subsection{Conservation Equations} 
\label{sec:sec21}

For a spherical cluster of massive stars that lose mass through their stellar winds, leading to a local volume rate of mass loading source $s(r)$ 
at intra-cluster radius $r$, mass conservation requires that the cluster density $\rho$ and radial outflow speed $v$ follow
\beq
\frac{d (\rho v r^2)}{dr} = r^2 s(r)
\, .
\label{eq:masscon}
\eeq
Defining a radially scaled cumulative mass function,
\beq
S(r) \equiv r^{-3} \int_0^r \, s(r') r'^2 \, dr'
\, ,
\label{eq:Sdef}
\eeq
the mass flux density then takes the form
\beq
\rho v = r \, S(r)
\, ,
\label{eq:rhov}
\eeq
while its spatial derivative is
\beq
\frac{d(\rho v)}{dr} = s - 2S \, .
\label{eq:drhov}
\eeq
For the standard simple case of a steady wind outflow with mass loss rate ${\dot M}$ fixed at the origin,
we have $s(r) = ({\dot M}/4 \pi r^2) \, \delta(r)$, giving $\rho v = {\dot M}/(4 \pi r^2)$.

For an energy-averaged terminal speed $V_w$ of the individual stellar winds, then upon collision with other winds,  
the average specific energy input to the intra-cluster medium is $V_w^2/2$.
For a cluster of total mass $M$ and size scale $w$, the ratio of cluster escape energy to wind energy is
\beq
g \equiv \frac{2 GM}{w V_w^2}  = 8.9 \times 10^{-4} \frac{M_5}{w_{pc} V_8^2}
\, ,
\label{eq:gscale}
\eeq
where the latter equality shows this is generally quite small for typical  cluster parameters, $M_5=M/(10^5 M_\odot)$,
$w_{pc}= w/pc)$ and $V_8 = V_w/(10^8$\,cm/s).
This indicates gravitational effects are negligible for clusters without a central black hole \citep[for that case, see, e.g.,][]{Silich08}.

For gas pressure $P$, adiabatic index $\gamma$, and specific enthalpy 
\beq
h \equiv \frac{\gamma}{\gamma -1} \, \frac{P}{\rho}
\label{eq:hdef}
\, ,
\eeq
the adiabatic\footnote{The analysis in section \ref{sec:pertcool} (see eqn.\ \ref{eq:epsLam}) indicates that radiative losses are generally a small correction to the adiabatic treatment assumed here..}
 conservation of energy thus takes the form
\beq
\frac{1}{r^2} \, \frac{d (\rho v r^2( v^2/2 + h))}{dr} = \frac{s V_w^2}{2}
\, .
\label{eq:eneq1}
\eeq
Upon integration and using eqn.\ (\ref{eq:rhov}), 
this reduces to a simple  Bernoulli relation between speed and enthalpy,
\beq
v^2 + 2 h = V_w^2
\, .
\label{eq:bern}
\eeq

Finally, the conservation of momentum takes the form
\beq
\rho v \frac{dv}{dr} = - \frac{dP}{dr} - s v
\, ,
\label{eq:momcons}
\eeq
where the last term represents an associated ``drag"  term stemming from mass addition at zero speed.
Using eqns. (\ref{eq:rhov})-(\ref{eq:hdef}),  (\ref{eq:bern}), and (\ref{eq:momcons}),
and defining the mass source ratio 
\beq 
\sigma(r) \equiv \frac{s(r)}{S(r)}
\, ,
\label{eq:sigdef}
\eeq
we obtain, through some extensive but straightforward algebra,
\beq
{
\frac{dv}{dr}= \frac{v}{r}\,  \frac{ 2 \gamma \sigma v^2  + (\gamma-1)(\vwsq- v^2)(\sigma-2)}{(\gamma-1)\vwsq - (\gamma + 1)v^2}
\, . 
}
\label{eq:eom}
\eeq
Our focus here will be on the most physically relevant case of a monatomic gas with $\gamma = 5/3$, 
for which eqn.\  (\ref{eq:eom}) becomes
\beq
\boxed{
\frac{dv}{dr}= \frac{v}{r}\,  \frac{ (4 \sigma+2) v^2  + (\sigma-2) \vwsq}{\vwsq -4 v^2}
\, . }
\label{eq:eomgg}
\eeq
The box highlighting emphasizes that this is a key equation for the analysis to follow.

In particular, in both eqns.\ (\ref{eq:eom}) and (\ref{eq:eomgg}), the vanishing of the denominator implies a critical speed,
\beq
v_c \equiv \sqrt{\frac{\gamma -1}{\gamma+1}} \vw = \frac{\vw}{2}
\,  ,
\label{eq:vcdef}
\eeq
where the latter equality takes $\gamma=5/3$.
The adiabatic sound speed $c_s \equiv \sqrt{\gamma P/\rho}$ is related to the specific enthalpy through $h = c_s^2/(\gamma-1)$. 
Applying the Bernoulli eqn.\ (\ref{eq:bern}) for this critical speed then shows that it equals the critical value of the sound speed, $v_c=c_{sc} $.
 
A smooth trans-critical solution also requires  that the numerators in (\ref{eq:eom}) and (\ref{eq:eomgg}) must vanish, 
which can occur at a critical radius $r_c$ ,
where 
\beq
\sigma (r_c)  = \frac{2}{\gamma+1}= \frac{3}{4}
\label{eq:rcdef}
\eeq
 for the monatomic case $\gamma=5/3$.
This critical radius can thus be physically identified as the {\em sonic} radius, where the flow goes from subsonic to supersonic.

The ability to identify {\it a priori} distinct values for the critical/sonic radius $r_c$ and the associated critical/sonic speed is an important,
novel feature of the notation and associated analysis developed here.
As demonstrated below, by facilitating a simple, direct numerical integration both inward and outward from this critical point, it allows one
to readily derive the full velocity solution $v(r)$ for almost {\em any} chosen mass deposition form $s(r)$.
 
\subsection{Power-law mass addition within a  fixed outer radius}
\label{sec:sec22}

An initial model by \citet{Canto00} simply assumed a {\em constant} mass addition $s$ within a cluster that has a {\em fixed outer radius}, beyond which it drops discontinuously to zero.
A follow on study by \citet{Rodriguez-Gonzalez07} generalized this to assuming the mass addition within this fixed outer radius follows a power-law form $s \sim r^p$.
For all such cases, this outer cluster radius now also serves as the critical/sonic radius $r_c$, with, however, now a steep, unbounded
critical point slope in velocity, $(dv/dr)_c \rightarrow \infty$.

As noted by \citet[][see their eqn.\ 15]{Canto00}, solutions for the  supersonic region $r>r_c$  can be derived from the Bernoulli eqn.\ (\ref{eq:bern}).
Applying the mass conservation eqn.\ (\ref{eq:masscon}), we find for an adiabatic outflow $c_s^2 \sim T \sim \rho^{\gamma-1} \sim 1/(v r^2)^{\gamma-1}$.
Using $c_{sc}^2/V_w^2 = (\gamma -1)/(\gamma +1) = 1/4$ for the standard case with $\gamma=5/3$,  the Bernoulli eqn.\ (\ref{eq:bern})  
 can then be recast into the form given by eqn.\ (15) of \citet{Canto00},
\beq
\frac{v}{ \vw} \left (1- \frac{v^2}{\vwsq}  \right )^{3/2} = \frac{3 \sqrt{3}}{16}  \, \frac {r_c^2}{r^2} 
\, ,
\label{eq:c15}
\eeq
which then can  be solved to obtain the velocity in the supersonic region, $r > r_c$ and $v > v_c = V_w/2$.

From the critical sonic speed at $r_c$, it is clear from eqn.\ (\ref{eq:c15}) that at large radii $r \rightarrow \infty$,
the cluster outflow speed must either approach zero or the wind speed $V_w$.
The latter case of supersonic outflow is the one with a vanishing asymptotic pressure, and so is preferred for cluster outflow 
into a low-pressure interstellar medium.
(See section \ref{sec:sec51} for further discussion of the role of interstellar medium pressure.)

The full solution requires matching the subsonic outflow in the mass-loaded cluster region with the supersonic 
outflow in the region beyond the outer cluster radius $r_c$.
For a mass source that follows a power-law variation in radius, $s \sim r^p$, we see that $\sigma = p+3$ is  spatially constant.
As first demonstrated by \citet{Canto00}, this allows one to separate the velocity and radial variations in eqn.\ (\ref{eq:eom}), 
which  \citet{Rodriguez-Gonzalez07}  show also applies for the more-general power-law case.
Upon integration and exponentiation to convert resulting logarithms back to radius and speed, 
one can derive a fully {\em analytic} forms for the velocity $v(r)$ in the subsonic region $r \le r_c$.
Models with $p<-1$ have divergent acceleration near the origin, which
\citet{Rodriguez-Gonzalez07} identify with models that have non-zero  speed at the origin,
but for our analysis here, we require $p>-1$.

\begin{figure}
\begin{center}
\includegraphics[scale=0.5]{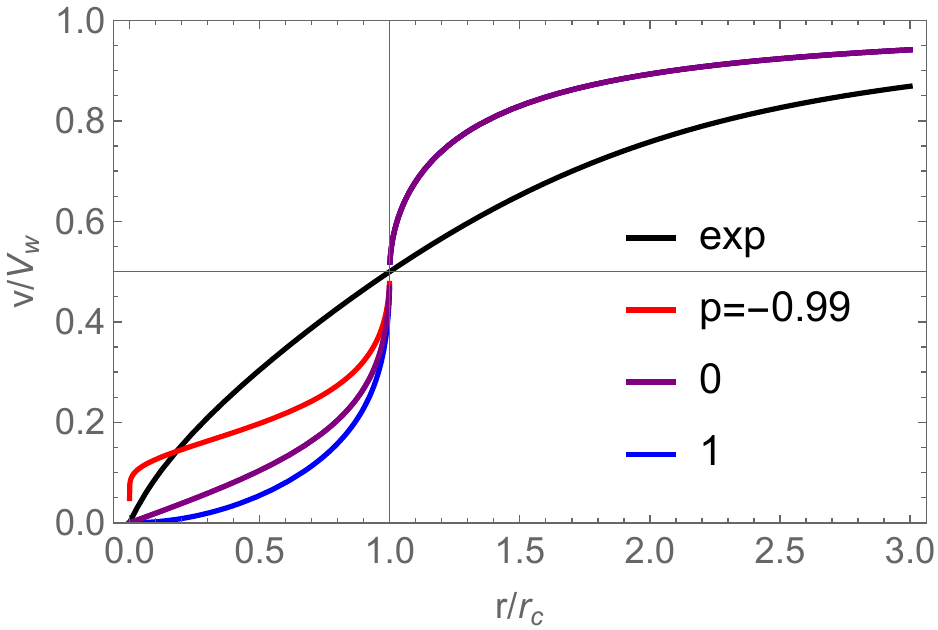}
\caption{For adiabatic expansion models with a fixed outer radius $r_c$, the colored curves plot outflow speed $v$ vs.\ radius $r$ using the solution for the labeled values of the mass-loading power index $p$.
Note that all models have divergent slope at the critical, outer radius $r_c$.
By comparison the black  curve shows the more gradual transonic outflow for the exponential deposition model suggested by \citet{Silich11},
but here using the adiabatic flow solution from eqn. (\ref{eq:eomgg}) that ignores effects of radiative cooling terms they considered.}
\label{fig:fig1}
\end{center}
\end{figure}

For the standard case of a monatomic ideal gas with  $\gamma=5/3$, 
the colored solid curves in the top panel of figure \ref{fig:fig1} compare the simple case with constant mass loading ($p=0$; purple curve) with
non-zero power index models $p=1$ (blue curve) and $p=-0.99$ (red curve).
All 3 have diverging slope at the critical radius, with a common outer (supersonic) solution given by eqn.\ (\ref{eq:c15}) (or eqn.\ (15)
from \citet{Canto00}).
The case $p=1$ with radially increasing mass loading has a shallower initial acceleration near the origin,
while the case $p=-0.99$ with radially {\em decreasing} $s(r)$ has a quite steep initial acceleration near the origin.

The black curve with a smoother transonic outflow is for a exponential mass-deposition model without a discrete outer boundary,
as we discuss next.

\subsection{Exponential deposition model without a fixed outer boundary}
\label{sec:exp}
\label{sec:sec23}

Instead of this simplified assumption of a sharp outer boundary, let us now consider models with a smoother drop off in stars and 
the associated mass deposition.
In particular, work by  \citet{Silich11} suggested a spatially unbounded mass deposition that follows an {\em exponential}
distribution of stars within a cluster, giving then for the mass deposition
\beq
s_e (r) =  \frac{{\dot M}}{8 \pi  w_e^3}  e^{-(r/w_e)} 
\, ,
\label{eq:se}
\eeq
where  ${\dot M}$ is the total mass loss rate, and
the cluster half-width $w_e$ is defined as the radius at which $s(w_e)/s_e(0) = 1/e$.
The associated ratio function $\sigma(r) \equiv s(r)/S(r)$ takes the form
\beq
\sigma_e (r) = 
\frac{(r/w_e)^3}{2 e^{r/w_e}-(r/w_e)^2-2 (r/w_e)-2}
\, .
\label{eq:sige}
\eeq
Since this is no longer spatially constant, it precludes the velocity-radius separation approach that allowed derivation of  analytic solutions
for above cases of power-law mass deposiition with a fixed outer radius.

But if we retain the assumption of purely adiabatic expansion, which ignores radiative cooling terms included in the analysis by \citet{Silich11} (see next subsection),  we can
derive the full critical solution by direct integration both  inward and outward from this critical/sonic point,
Specifically, applying this requirement $\sigma_e(r_{ce})=3/4$ for this case of an exponential mass deposition, we find $r_{ce} = 4.06 w_e$,
where $v(r_{ce}) = v_c = V_w/2$.

In terms of the radius scaled by the critical radius, the black curve in 
figure \ref{fig:fig1} compares velocity variation for this exponential deposition model to the previous power-law depositions, showing now that it has a much more gradual acceleration through the sonic radius, while still approaching a terminal speed set by the mass-weighted wind speed $V_w$ at very large radii.

\subsection{Radiative cooling as a perturbative effect}
\label{sec:pertcool}
\label{sec:sec24}

For this exponential mass deposition case, Silich et al. (2011) derived results using a much more challenging and numerically delicate `shooting method', based on conservation equations that include radiative cooling effects that in principle can override the previous assumption of purely adiabatic expansion.
The analysis in this section now shows, however, that such radiative cooling generally represents only a small perturbative effect to adiabatic expansion solutions.

Adding radiative cooling to the energy equation (\ref{eq:eneq1}) gives
\beq
\frac{1}{r^2} \, \frac{d(\rho v r^2(v^2/2+h))}{dr} = \frac{s V_w^2}{2} - \rho^2  {\bar \Lambda}
\, ,
\label{eq:dvdrlam}
\eeq
where the term with $ {\bar \Lambda} \equiv  \Lambda/(\mu_e \, \mu_p)$ accounts for radiative cooling,
with $\Lambda (T)$  the usual optically thin cooling function \citep[e.g.,][]{Schure09}.
For standard hydrogen mass fraction $X=0.72$, the mean molecular weights
per proton and per electron are given in terms of the proton mass $m_p$ by
respectively $\mu_p = m_p/X  $ and  $\mu_e = 2 m_p/(1+X)$.

Integration of eqn. (\ref{eq:dvdrlam}) now gives a specific enthalpy set by
\beq
2 h =V_w^2 - V_\Lambda^2 -  v^2
\, ,
\label{eq:hlamdef}
\eeq
where the radiative effect is written in terms of a specific  energy that can be cast in terms of
the square of a `cooling speed',
\beq
V_\Lambda^2 (r) \equiv \frac{2 \int_0^r r'^2 \rho^2 {\bar \Lambda}(T)\, dr'}{\rho v r^2}
\, .
\label{eq:VLamdef}
\eeq
Recapitulating the procedure leading to  equation (\ref{eq:eomgg}), we find that,
after some manipulation, the velocity gradient equation now  takes the form,
\beq
\frac{dv}{dr}
= \frac{v}{r} ~ \frac{ ( 4 \sigma +2) v^2 +   (\sigma -2) (V_w^2 - V_\Lambda^2)
- 2r \frac{d V_\Lambda^2}{dr}  }
{  V_w^2 - V_\Lambda^2 - 4 v^2 }
\, .
\label{eq:dvdrLam}
\eeq
The appearance now of the cooling speed $V_\Lambda$ in both the denominator and numerator of the right-hand-side
implies a potential change in respectively the critical (sonic) speed $v_c$ and the location of the critical/sonic radius $r_c$.

Since this cooling speed correction depends itself on the full velocity solution, one in general can no longer identify {\it a priori}
the critical  radius and critical speed; thus instead of the above procedure of simply integrating inward and outward from the critical
radius, a full general solution of eqn.\ (\ref{eq:dvdrlam}) can  now require an elaborate `shooting method' approach
to evolve a model from some initial subsonic speed past a flow critical sonic point, where the denominator
vanishes; this was the approach taken by \citet{Silich11}.

To explore the relative importance of such cooling, let's examine the overall scaling of $V_\Lambda$.
Scaling the speed $v$ by the wind speed $V_w$, the radius $r$ by the cluster half-width $w$, and the density $\rho$ by
${\dot M}/(4 \pi w^2 V_w)$,  we can see that the overall scale $\epsilon_\Lambda$ of the energy ratio $V_\Lambda^2/V_w^2 $ is given by
\beq
\boxed{
\epsilon_\Lambda =
 \frac{{\bar \Lambda}_o {\dot M} }{2 \pi w V_w^4} 
 \approx  2.1 \times 10^{-4} \frac{{\dot M}_{-3}}{w_{pc} V_8^4} 
 }
\, ,
\label{eq:epsLam}
\eeq
where\footnote{The inverse ratio $\chi \equiv 1/\epsilon_\Lambda$ has the same basic scaling proposed by \citet{Stevens92} for the
relative importance of expansive versus radiative cooling in colliding wind binary shocks.}
 the latter evaluation assumes a Hydrogen mass fraction $X=0.72$, standard cosmic abundances\footnote{The radiative cooling function $\Lambda$ scales roughly with metalicity $Z$, so could be proportionally lower or higher in metal poor or metal rich clusters.},  and a cooling function
$\Lambda_o  \approx 3 \times 10^{-23} {\rm erg \, cm^3/s}$, appropriate
for  typical wind thermalization temperatures of order  a few times $ 10^7$\ K \citep{Schure09}.
Here $w_{pc} \equiv w/$pc,   $V_8 \equiv V_w/(10^8$cm/s), and ${\dot M}_{-3} \equiv {\dot M}/10^{-3} (M_\odot$/yr)
are typical scalings for a parsec sized cluster fed by $\sim$100 WR stars with individual winds of mass loss rate $\sim 10^{-5} M_\odot$/yr and
terminal speed $V_\infty \approx $1000\,km/s.

\begin{figure}
\begin{center}
\includegraphics[scale=0.55]{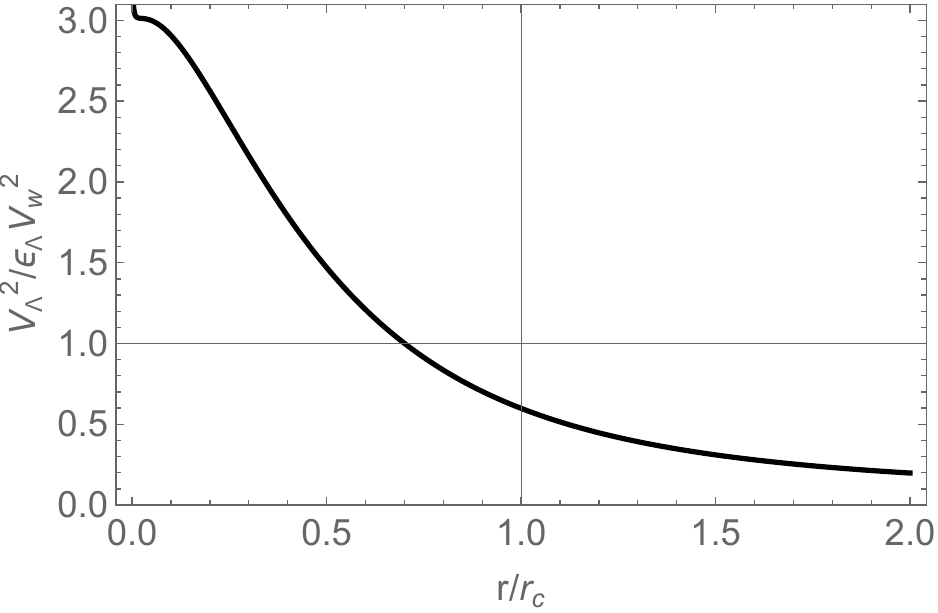}
\caption{Using the adiabatic solution for exponential mass deposition, the radial variation of the square of the cooling speed given by eqn.\ (\ref{eq:VLamdef}), scaled by the square of the wind speed $V_w^2$ times the scale factor $\epsilon_\Lambda$ given by eqn.\ (\ref{eq:epsLam}).
}
\label{fig:fig2}
\end{center}
\end{figure}

The fact that, for these standard parameters,  $\epsilon_\Lambda \ll 1 $ indicates cooling
should generally have only a small, perturbative effect, allowing one then to evaluate the radial variation of  $V_\Lambda$ by applying  the adiabatic solution in an explicit integration of (\ref{eq:VLamdef}).
For the above case of exponential mass deposition, figure \ref{fig:fig2} shows the scaled energy ratio $V_\Lambda^2/\epsilon_\Lambda V_w^2$  is of order unity, declining from about 3 near the cluster center, to below unity near the flow critical radius.
Combined with the scale factor  $\epsilon_\Lambda  \ll 1$, the implication is that the correction to the  simple adiabatic solution will be generally small, with nearly the same critical radius and critical speed.

For comparison, the full radiative cooling models computed  by \citet{Silich11} using the shooting method assume $w_{pc}=V_8=1$, but with
3 assumed wind luminosities, corresponding to 3 cluster mass loss rates (see their Table 1).
Their most extreme model C has $M_{-3} = 947$, representing now an $\epsilon_\Lambda = 0.20$ that, while approaching order unity, is still modest; their models A and B have respectively a factor  ten and hundred lower mass loss rates, implying $\epsilon_\Lambda \ll 1$ and so 
negligible cooling effects.
 
The upper left panel of their Fig. 2 compares their computed velocity laws, with models A and B  indeed showing nearly identical forms that asymptote at the wind terminal speed $V_w = 1000$\ km//s, as expected for their effectively adiabatic acceleration.
Model C shows a modest 10\% reduction in terminal speed, quite consistent with its $\epsilon_\Lambda \approx 0.2$, and the result from figure \ref{fig:fig2} that shows the scaled cooling speed asymptotes 
below 0.5 at large radii.

Within the usual scenario that such cluster outflows are fed mainly by massive Wolf-Rayet (WR) stars with typical individual mass loss rates $\sim 10^{-5} M_\odot$/yr, we see that these 3 models would require clusters ranging from $10^3$ to $10^5$ such WR stars, much more than the $\sim$100 such stars taken here as typical of massive young clusters.
Nonetheless, we see that even for the most extreme model C, radiative cooling is still a modest correction to the flow dynamics.
The upshot is that one should quite generally be able to derive solutions from sonic point integration of eqn.\ (\ref{eq:eomgg}),
allowing for much greater freedom to explore a range of mass deposition models, as we now demonstrate.

\section{Alternate Forms for Mass Deposition}
\label{sec:sec3}

One issue for the above exponential model is that it is based on a stellar mass distribution
that has a non-zero gradient at the origin.
For a stellar cluster that is bound entirely by the {\em self}-gravity of the ensemble of stars
(without, e.g., a central mass like a massive black hole), the fact that the mass $M(r)$ within a given
radius vanishes for $r \rightarrow 0$ implies that the gradient in stellar mass density should also
vanish, $d\rho/dr |_{r \rightarrow 0}  \rightarrow 0$. For the usual picture that  the mass deposition from stellar winds
scales in proportion to the stellar density, this also requires $ds/dr |_{r \rightarrow 0} \rightarrow 0$ at the cluster center.

This can be satisfied by even functions for which the mass deposition scales with $r^2$ (or indeed all even powers of $r$).
The next  subsections detail  3 such models.

\begin{figure}
\begin{center}
\includegraphics[scale=0.5]{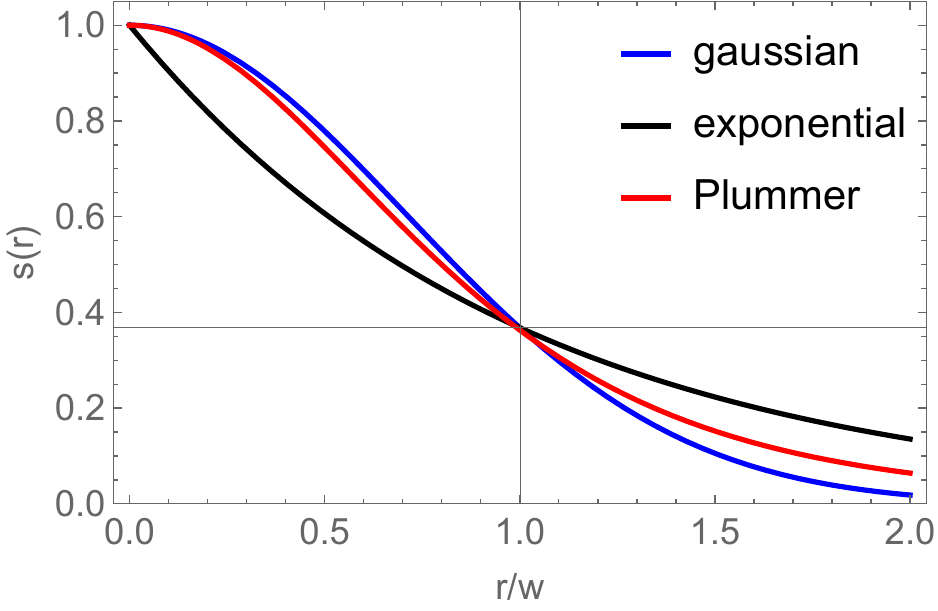}
\includegraphics[scale=0.5]{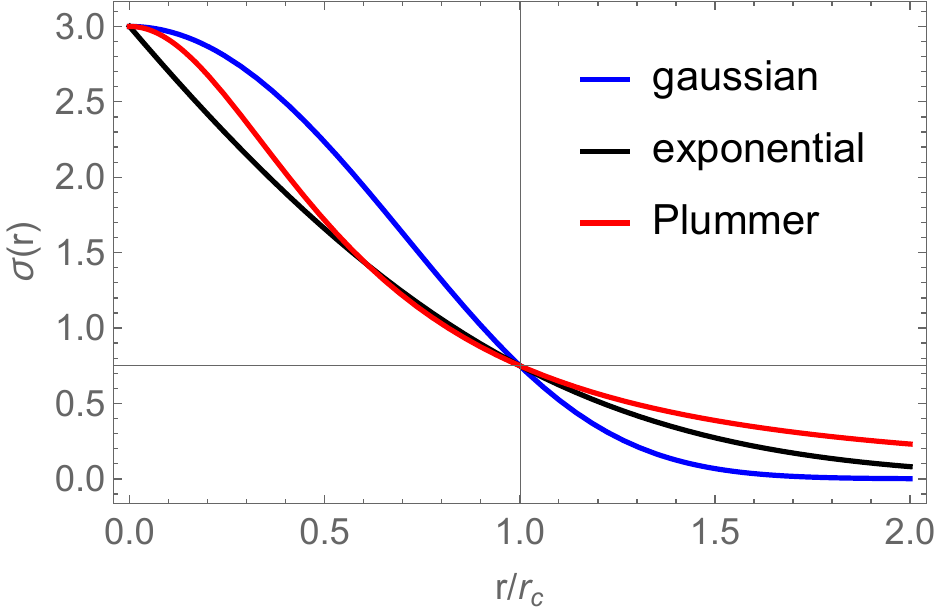}
\includegraphics[scale=0.5]{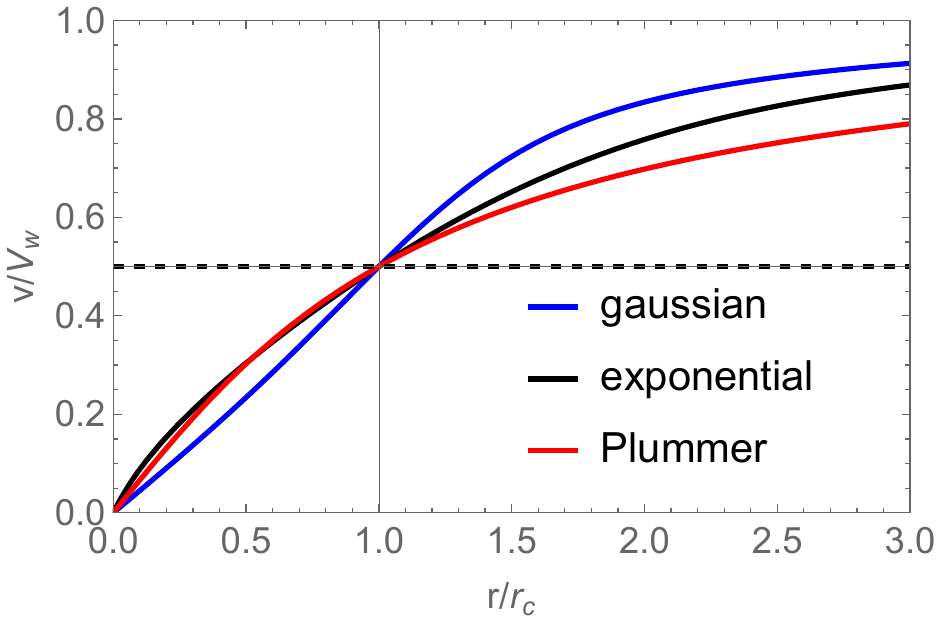}
\caption{Top: Three forms for assumed mass deposition $s(r)$ plotted vs. radius scaled by the half-width $w$ at which the deposition has been reduced by a factor $1/e$ (denoted by the horizontal line). 
Middle: Associated mass deposition ratios $\sigma(r)$ plotted vs. radius scaled by the critical radius $r_c$, at which $\sigma(r_c) = 3/4$ (again denoted by the horizontal line).
Bottom: Associated solutions for radial variation of velocity $v(r)$, again with radius scaled by $r_c$.}
\label{fig:fig3}
\end{center}
\end{figure}

\subsection{Gaussian model for  deposition}
\label{sec:sec31}

As a first example, let us assume the mass deposition follows a canonical gaussian form,
\beq
s_g(r) \equiv \frac{{\dot M}}{\pi^{3/2} w_g^3}  e^{-(r/w_g)^2} 
\, ,
\label{eq:sg}
\eeq
where  ${\dot M}$ is the total mass loss rate, and
the cluster half-width $w_g$ is again defined as the radius at which $s_g(w_g)/s(0) = 1/e$.
\newcommand{\erf}{\ensuremath{\operatorname{erf}}}
The  cumulative mass function takes the form
\beq
S_g(r) =  
r^{-3} ( {\dot M} \erf (r/w_g)/4\pi - r w_g^2 s(r) /2 )
\, .
\label{eq:Sg}
\eeq
The  associated deposition ratio $\sigma_g(r) = s_g(r)/S_g(r)$ is then given by the ratio of eqns. (\ref{eq:sg}) and (\ref{eq:Sg}).

Applying the critical point condition $\sigma_g(r_{cg}) = 3/4$ gives $r_{cg} = 1.67 w_g$.
Integration inward and outward from this critical sonic point gives the full velocity solution $v(r)$ plotted 
as the blue curve in the bottom panel of figure \ref{fig:fig3}.

\subsection{Mass deposition based on the empirical cluster model by Plummer}
\label{sec:sec32}

Another prominent  distribution with zero gradient at the origin is the form derived by \citet{Plummer11}
for the mass density distribution of globular clusters, which we use here to define a mass deposition of the form
\beq
s_p (r) \equiv 
\frac{s_{po}}{(1 + r^2/2w_p^2)^{5/2}}
\, ,
\label{eq:spdef}
\eeq
where $s_{po}  \equiv 3 {\dot M}/8\sqrt{2}\pi w_p^3$ and the half-width $w_p$ again gives $s_p(w_p) \approx s_{po}/e$.
Carrying out the integration in eqn.\ (\ref{eq:Sdef}) for this new mass source,  we obtain the 
cumulative mass function,
\beq
S_p(r) = \frac{ s_{po}}{3(1 + r^2/2w_p^2)^{3/2}}
\, ,
\label{eq:CapSpdef}
\eeq
giving then a mass deposition ratio,
\beq
\sigma_p (r)  \equiv \frac{s_p (r)}{S_p (r)} = \frac{3}{1 + r^2/2w_p^2}
\, .
\label{eq:sigpdef}
\eeq
From the requirement $\sigma_p (r_{cp})= 3/4$, we find for this case $r_{cp}= 2.45 w_p$.

Figure \ref{fig:fig3} compares the radial variations of these three labeled forms of mass deposition $s$ (top: in units of the respective half-widths
$w$) and their associated deposition ratio $\sigma$ (middle: in units of their respective critical radii $r_c$.)
For all three models,
straightforward integration of eqn.\ (\ref{eq:eomgg}) inward and outward from their respective critical/sonic points gives the full radial variation of the flow velocity, as plotted in the bottom panel of figure \ref{fig:fig3}.

Overall, when cast in terms of the radius scaled by the associated critical radius, the velocity variations of all 3 models are quite similar,
anchored by the need to start near zero at the origin, pass through the critical/sonic speed $v_c = c_{sc} = V_w/2$ at $r_c$, and
asymptote to the mass-weighted stellar wind speed $V_w$ at large distances.
The upshot is that, within this base class of models, the characteristics of wind outflows are quite robust, and insensitive to details of
the mass deposition model.

\begin{figure}
\begin{center}
\includegraphics[scale=0.6]{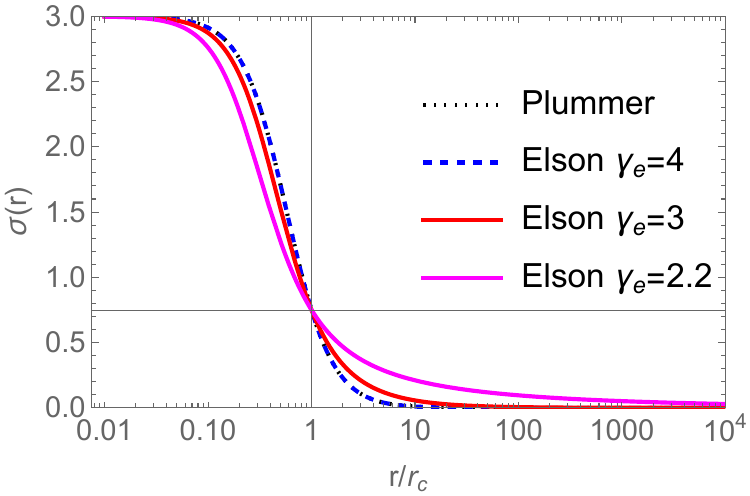}
\includegraphics[scale=0.6]{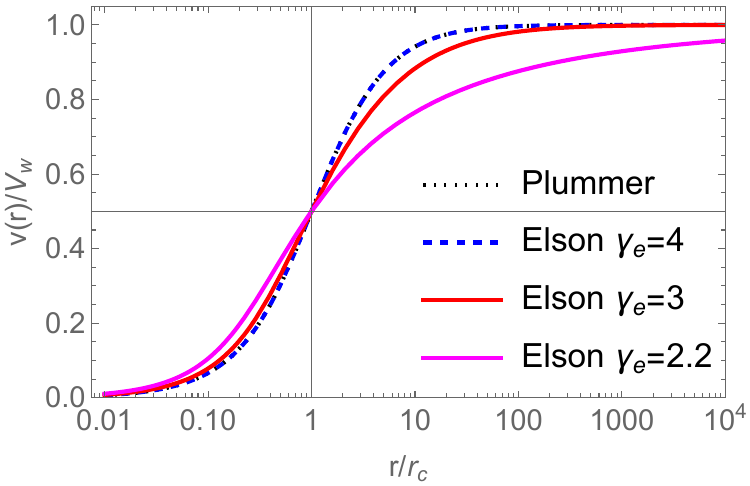}
\caption{Top: Radial variations of the mass deposition ratio $\sigma$ plotted against the radius on a log scale, comparing variations for the Elson models with labeled exponent parameters $\gamma_e$, along with the Plummer model, which is equivalent to the Elson model with $\gamma_e=4$.
Note that the $\gamma_e = 2.2$  case declines only slowly to zero.
Bottom: Associated velocity variations,  showing that this $\gamma_e = 2.2$ case now also only slowly approaches the terminal value $V_w$.
 }
\label{fig:fig4}
\end{center}
\end{figure}

\subsection{Mass deposition based on the empirical cluster models by Elson}
\label{sec:sec33}

In their extensive empirical study of the density distribution of young stellar clusters in the Large Magellanic Cloud (LMC), 
 \citet{Elson87} proposed (see their eqn. (13a)) a generalized form of the  \citet{Plummer11} power-law in eqn.\ (\ref{eq:spdef}),
 but with the 5/2 exponent replaced by the parametrized expression $(\gamma_e + 1)/2$, wherein finite mass normalization
 requires $\gamma_e > 2$.

Accordingly, let us now consider mass depositions of the form
\beq
s_l (r,\gamma_e) \equiv 
\frac{s_{lo}}{(1 + c_{\gamma e} r^2/w_l^2)^{(\gamma_{e}+1)/2}}
\, ,
\label{eq:sldef}
\eeq
where defining 
\beq
c_{\gamma e} \equiv e^{2/(\gamma_e+1)} - 1
\, 
\label{eq:cgdef}
\eeq
ensures that $s_l(w_l,\gamma_e) = s_{lo}/e$.
Carrying out the integration in eqn.\ (\ref{eq:Sdef}) for this mass source,  we obtain 
the  cumulative mass function,
\beq
S_l(r) =
\frac{s_{po}}{3} \, _2F_1\left(\frac{3}{2},\frac{\gamma_e
   +1}{2};\frac{5}{2};-c_{\gamma e} r^2/w_l^2\right) 
 \, ,
\label{eq:CapSldef}
\eeq
where $ _2F_1$ is a hypergeometric function.
This gives for
the mass deposition ratio,
\beq
\sigma_l (r,\gamma_e)  
= 
\frac{3}{\left(\frac{c_{\gamma e} r^2}{w_l^2}+1\right) \,
   _2F_1\left(1,2-\frac{\gamma_e
   }{2};\frac{5}{2};-\frac{c_{\gamma e} r^2}{w_l^2}\right)}\, .
\label{eq:sigldef}
\eeq

Now plotting variations out to large radii on a log scale, the upper panel of  figure \ref{fig:fig4} shows that lowering the power index $\gamma_e$ close to its limiting value $\gamma_e \rightarrow 2$ leads to a much more extended mass deposition, with $\sigma _l$ declining only slowly (almost logarithmically) in radius.
The lower panel shows that this leads to a much slower flow acceleration, with the velocity still nearly 10\% below its terminal value  even at an extreme radius $r \approx 10,000 \, r_c$.

\section{Associated variations}
\label{sec:sec4}

\subsection{Variations in temperature and density}
\label{sec:sec41}

The above solutions for scaled flow speed $v(r)$ can be used to derive 
associated variations in temperature and density.
Noting again that $h \sim T$, we see from the Bernoulli eqn.\ (\ref{eq:bern}) that 
the temperature is given by
\beq
\frac{T}{T_o} = 1 - \frac{v^2}{\vwsq}
\, ,
\label{eq:Todef}
\eeq
Since the speed $v$ vanishes at the origin, we see from the Bernoulli relation (\ref{eq:bern}) that
the central temperature is given by \citep[cf.][]{Canto00, Silich04}
\beq
T_o = \frac{\mu V_w^2}{5 k} = 15 \, {\rm MK} \, V_8^2
\, .
\label{eq:To}
\eeq
The latter equality again uses $V_8 \equiv V_w/(10^8 \, {\rm cm/s})$, and assumes a standard molecular weight $\mu = 0.62 m_p$ appropriate to a fully ionized
mixture
 with hydrogen mass fraction $X=0.72$.

Using eqns.\ (\ref{eq:rhov}) and (\ref{eq:eom}), we find for the central density  
\beqa
\rho_o &=& \lim_{r \rightarrow 0} \left ( \frac{r S(r)}{v(r)} \right ) 
\nonumber \\
&\approx& \frac{s(0) r_c}{3 v_c}
\nonumber \\
 &\approx& 0.12 \, \frac{{\dot M}}{ w^2 V_w} \, \frac{r_c}{w}
 \nonumber \\
 &\approx&  7.8  \times 10^{-24} {\rm \frac{g}{cm^3} }\, \frac{{\dot M}_{-3}}{w_{pc}^2 V_8} \, \frac{r_c}{w}
\, ,
\label{eq:rhoo}
\eeqa
where the numerical evaluations here are for the gaussian mass deposition model.
For the other  depositions, the different normalizations would give slightly different numerical values for $s(0)$,
but we do explicitly account for differences in  the critical radius to $r_c/w$.
The constant deposition model by \citet{Canto00} gives a similar parameter dependence, but with a order-unity 
different numerical factor \citep[see also eqn.\ (1) of][]{Silich04}.

\begin{figure}
\begin{center}
\includegraphics[scale=0.5]{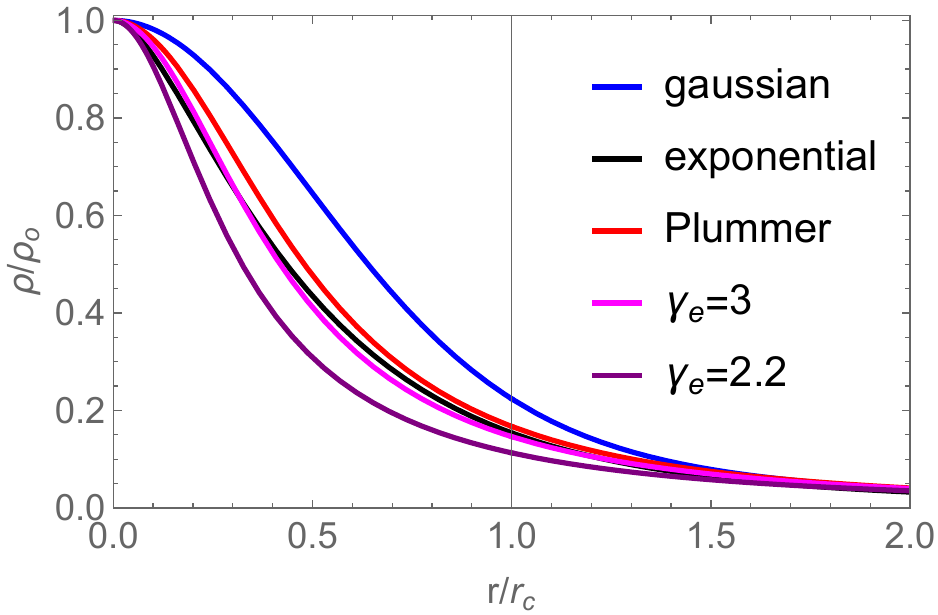}
\includegraphics[scale=0.5]{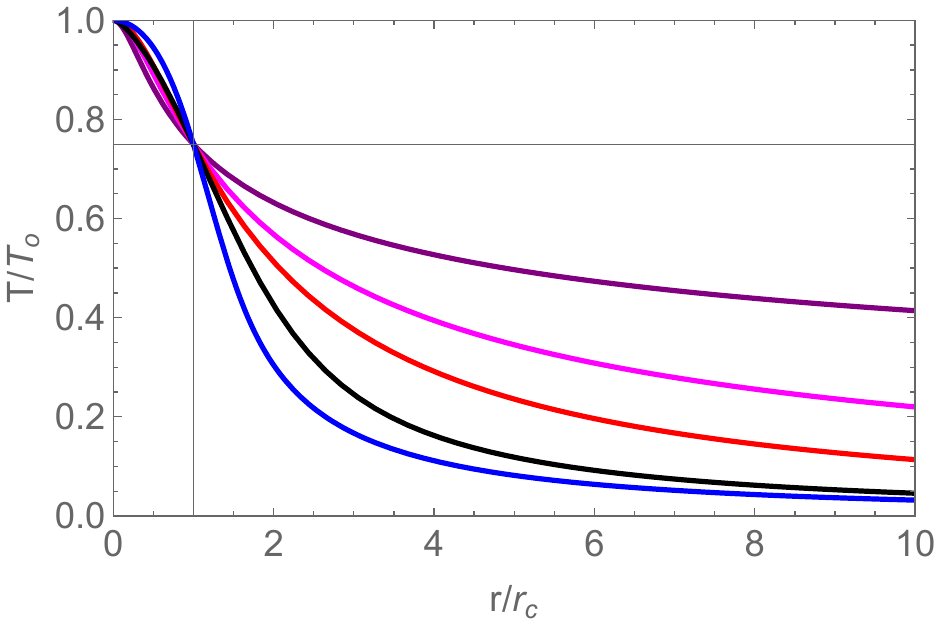}
\caption{For the various labeled models, the  variation of density (top) and temperature (bottom), scaled by their values at the cluster center, 
plotted versus radius scaled by the critical radius for each model, $r/r_c$.}
\label{fig:fig5}
\end{center}
\end{figure}

The radial variation of density $\rho(r)$ away from the origin can then be obtained from eqn.\ (\ref{eq:rhov}).
Figure \ref{fig:fig5} compares the radial variation of  $\rho/\rho_o$  and $T/T_o$ vs. $r/r_c$ for the five labeled models
derived above.

\subsection{Radiative losses}
\label{sec:radcool}
\label{sec:sec42}

With these results for density and temperature, let us now derive scalings for
the associated radiative losses.
For the central temperature and density,  the radiative energy lost per unit volume is
\beq
{\dot q }_o \equiv {\dot q }(\rho_o,T_o) 
= {\bar \Lambda}_o \rho_o^2
\, ,
\label{eq:qdotodef}
\eeq
where the evaluation of ${\bar \Lambda_o}$ is given following eqn.\ (\ref{eq:epsLam}).
Applying the scaling (\ref{eq:rhoo}) for $\rho_o$, we find 
\beq
{\dot q}_o= 4.1 \times 10^{-22} {\rm \frac{erg}{cm^3 \, s} } \, \left ( \frac{{\dot M}_{-3}}{w_{pc}^2 V_8} \right )^2  \left ( \frac{r_c}{w} \right )^2
\, .
\label{eq:qdotoval}
\eeq
We can define an associated scaling for the radiative luminosity as
\beqa
L_o  &\equiv& 4 \pi w^3 {\dot q}_o /3 
\nonumber \\
&=& 5.0 \times 10^{34} {\rm \frac{erg}{ s} } \, \frac{{\dot M}_{-3}^2 }{ w_{pc} V_8^2} \left ( \frac{r_c}{w} \right )^2
\nonumber \\
&\approx& 13 \, L_\odot \, \frac{{\dot M}_{-3}^2 }{ w_{pc} V_8^2} \left ( \frac{r_c}{w} \right )^2
\, .
\label{eq:Lo}
\eeqa
By comparison, the terminal value of the wind kinetic energy luminosity is gives by
\beq
L_{\rm ke} \equiv {\dot M} V_w^2/2  = 8.1 \times 10^4 L_\odot \, {\dot M}_{-3}  V_8^2
\, ,
\label{eq:Lkedef}
\eeq
from which we find that the ratio of radiative to kinetic luminosity scales with the cooling parameter defined in eqn.\ (\ref{eq:epsLam}),
\beq
\frac{L_o}{L_{\rm ke}} 
\approx  
1.6 \times 10^{-4} \frac{{\dot M}_{-3}}{w_{pc} V_8^4} \left ( \frac{r_c}{w} \right )^2
\approx 0.75 \, \epsilon_\Lambda \left ( \frac{r_c}{w} \right )^2
\, ,
\label{eq:LobLke}
\eeq
providing another characterization of the generally small effect of radiative losses, as was discussed in section \ref{sec:pertcool}.

\subsection{X-ray luminosity and surface brightness}
\label{sec:sec43}

From these variations in density and temperature, we can also derive
estimated scalings for the emission at  X-ray energies.
For X-rays above a given energy threshold $kT_x$, analysis by \citet{Uddoula14} (see their eqn.\ (26)) shows
that the volumetric emission above this energy threshold can be approximated by
\beq
{\dot q} (\rho,T,T_x) \approx  {\dot q}_o \left( \frac{\rho}{\rho_o} \right )^2 e^{-T_x/T} 
, .
\label{eq:qdotdef}
\eeq
For a given $T_x$, the associated X-ray surface brightness as a function of offset $p \equiv \sqrt{r^2 - z^2}$ from the center can be computed from integration along the depth coordinate $z$,
\beq
I_x (p,T_x) =  {\dot q}_o 
 \int_{-\infty}^\infty  \left ( \frac{\rho(r)}{\rho_o} \right ) ^2 e^{-T_x/T(r)}  dz
\, 
\label{eq:Ixdef}
\eeq
where within the integrand, the radius varies as $r = \sqrt{p^2 +z^2}$ for fixed offset $p$.
The associated X-ray luminosity is given by integration over this offset,
\beq
L_x(T_x) = 2 \pi \int_0^\infty I_x(p,T_x) \, p \, dp 
\, .
\label{eq:Lxpint}
\eeq

\begin{figure}
\begin{center}
\includegraphics[scale=0.5]{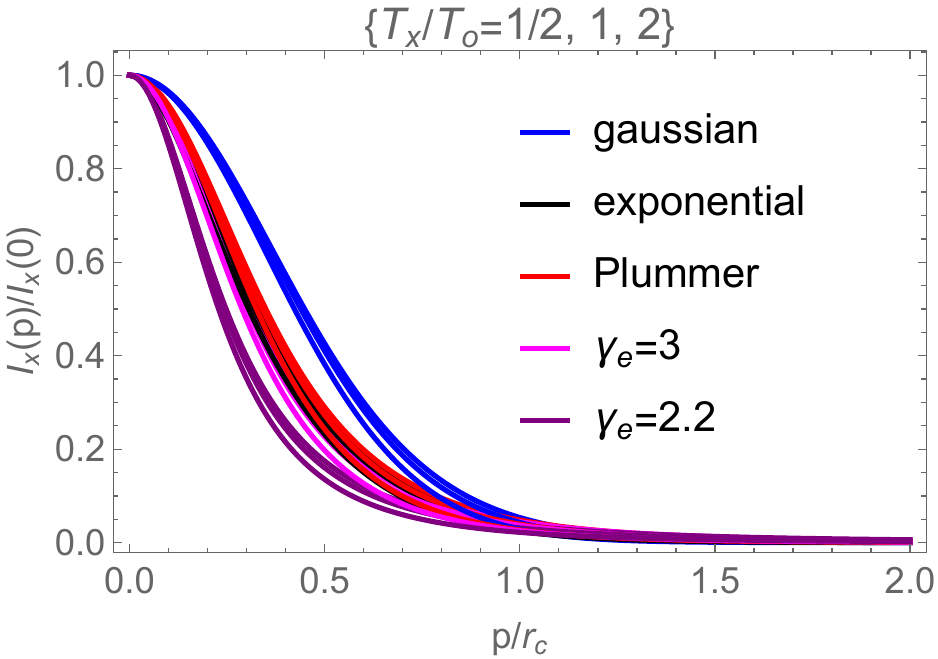}
\includegraphics[scale=0.5]{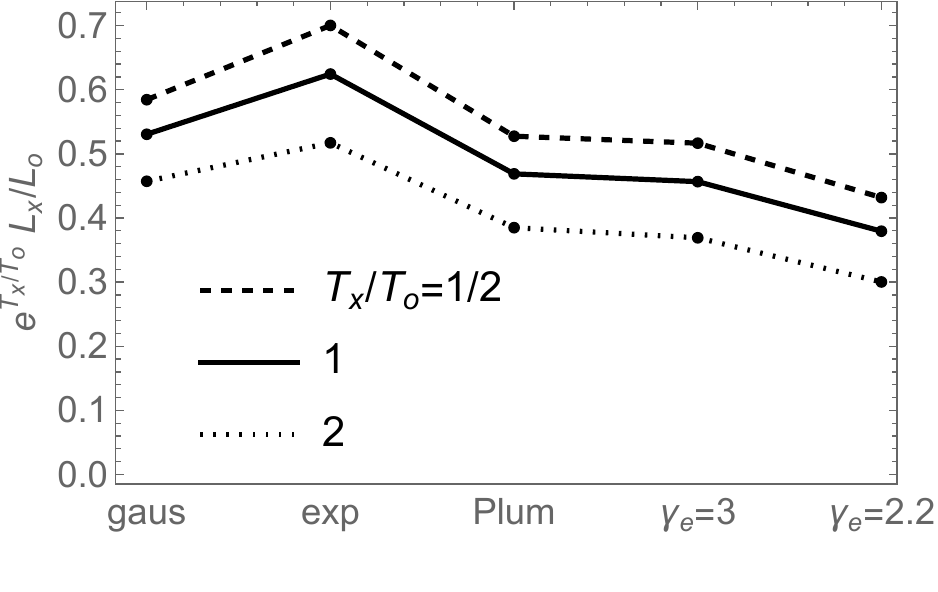}
\caption{Top: For the 3 cases  $T_x/T_o = 1/2, 1, 2 $, an overplot of surface brightness normalized by its central ($p=0$) value, 
plotted versus offset scaled by critical radius, $p/r_c$ for all 5 of the labeled mass deposition models.
Bottom: Associated integrated X-ray luminosities scaled by $e^{T_x/T_o}$ for the 3 $T_x$ cases and normalized by radiative luminosity $L_o$ given by eqn.\ (\ref{eq:Lo}).
The abscissa  labels  represent respectively the gaussian, exponential, Plummer, and Elson models (with $\gamma_e = 3$ and 2.2) for mass deposition.}
\label{fig:fig6}
\end{center}
\end{figure}

For the 3 cases  $T_x/T_o = $1/2, 1, and 2, the top panel of figure \ref{fig:fig6} overplots the spatial variation of $I_x(p,T_x)/I_x(0,T_x)$, the X-ray surface brightness normalized by that for the central ray with $p=0$.
This shows  that, when cast in terms of the $r/r_c$ and normalized by the central intensity $I_x(0,T_x)$, the overall spatial variation is quite similar for all 5 models;
indeed, when scaled by the factor $e^{T_x/T_o}$,  it is also similar for all 3 values of $T_x/T_o$.

The lower panel of figure \ref{fig:fig6} shows the integrated X-ray luminosity $L_x$  adjusted by the exponential factor as $e^{T_x/T_o}$ for the 3 $T_x$ cases, set in  units of the base radiative luminosity $L_o $ in eqn.\  (\ref{eq:Lo}); the horizontal axis labels represent respectively the gaussian, exponential, Plummer, and Elson models (with $\gamma_e = 3$ and 2.2).
When scaled in this way, the predicted X-ray luminosities have only a narrow range values, from about 0.7 to 0.4.
The upshot is that the overall scaling with cluster flow parameters is set by scaling for $L_o$ given in eqn.\ (\ref{eq:Lo}), modified by the factor $e^{-T_x/T_o}$ for different X-ray energy bands parameterized by $T_x$.

As noted already in the Introduction, observed X-ray luminosities are generally lower than theoretical
predictions from previous models based on fully thermalized winds \citep{Lopez11,Rosen14,Gupta18,Lancaster21}.
But the analytic scalings derived here provide a useful way to further quantify and analyze this discrepancy for a wide range of clusters with varying parameters and even mass deposition models.

\section{Extensions}
\label{sec:sec5}

\subsection{Interaction with the ISM}
\label{sec:sec51}

If the interstellar medium (ISM) around the cluster has a low pressure, the supersonic cluster outflow will
terminate in a reverse shock of a wind-blown bubble complex, as discussed by \citet{Weaver77}.
For ISM pressure $P_{\rm ISM}$, we can define a radius at which this balances the terminal wind 
momentum,
\beq
r_b = \sqrt{\frac{{\dot M} V_w}{4 \pi P_{\rm ISM}}}
\, .
\label{eq:rbdef}
\eeq
But the blue curve in figure \ref{fig:fig7}b shows that, for the typical case of the gaussian deposition model, the momentum flux density peaks near the cluster size  $w$.
Setting  $r_b = w$, we can thus use (\ref{eq:rbdef}) to estimate a  maximum pressure that can still 
allow a supersonic outflow (cf. eqn.\  24 of \citet{Canto00}),
\beq
\frac{P_{max}}{k} =( n T ) _{max}   \approx  \frac{{\dot M} V_w}{4 \pi w^2 k} = 3.8 \times 10^8 {\rm \frac{K}{cm^3}}
\frac{{\dot M}_{-3} V_8}{w_{pc}^2}
\, .
\label{eq:Pmaxdef}
\eeq
For ISM pressures above this maximum, we expect that cluster outflow will revert to subsonic ``breeze".
But the last evaluation in (\ref{eq:Pmaxdef}) indicates that 
this would require a quite high ISM pressure.
Even taking a lower mass loss rate ${\dot M_{-3}} = 0.1$ in a more extended cluster $w_{pc} = 3$ ($w \approx 3$\,pc), 
a subsonic breeze would require an ISM pressure that is $> 1000$ times a typical value in the galaxy,  
$(nT)_{\rm ISM} \approx 3000$\,K/cm$^3$ \citep{McKee77}.

Much higher ISM pressures can occur in starburst galaxies \citep{Westmoquette14}, or in
HII regions.
For such high-pressure cases, we can recast ({\ref{eq:Pmaxdef}) 
in terms of an associated maximum cluster size to still have a supersonic outflow,
\beq
w_{max}
\approx 20
  \, {\rm pc} \left ( \frac{{\dot M}_{-3} V_8 }{(nT)_6 } \right )^{1/2} 
\, ,
\label{eq:wmax}
\eeq
where $(nT)_6 \equiv nT/(10^6 {\rm K/cm^3})$. The upshot is that, except for the most extended, less-massive clusters embedded
in such a high-pressure ISM, we can expect the cluster outflow to be supersonic and terminate in a shock.

Note however that \citet{Pascale24} claim to have discovered a young super star cluster with a highly pressurized intra-cluster environment from gravitationally trapped stellar ejecta.

\begin{figure}
\begin{center}
\includegraphics[scale=0.5]{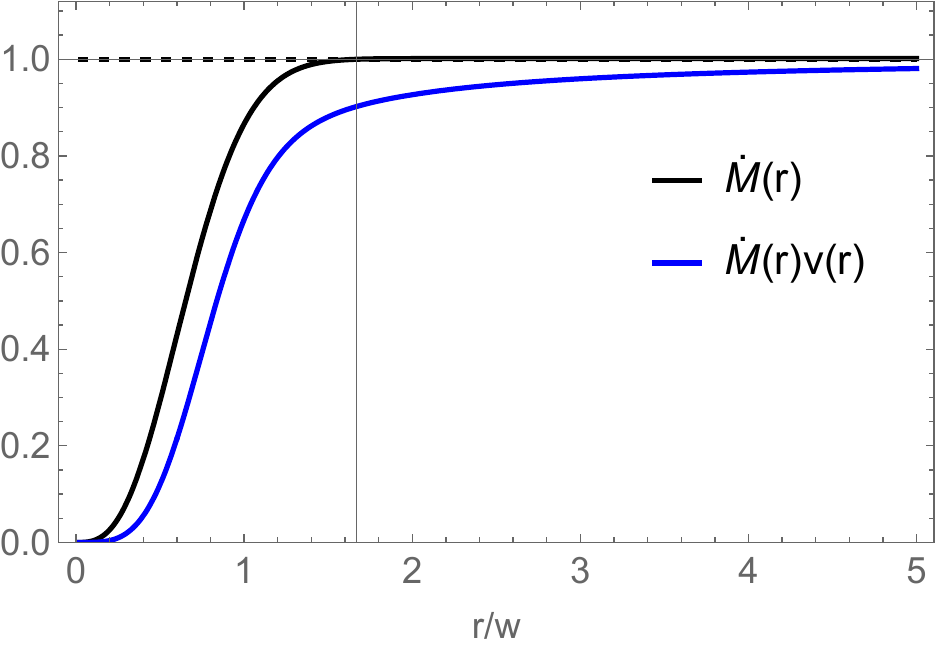}
\includegraphics[scale=0.5]{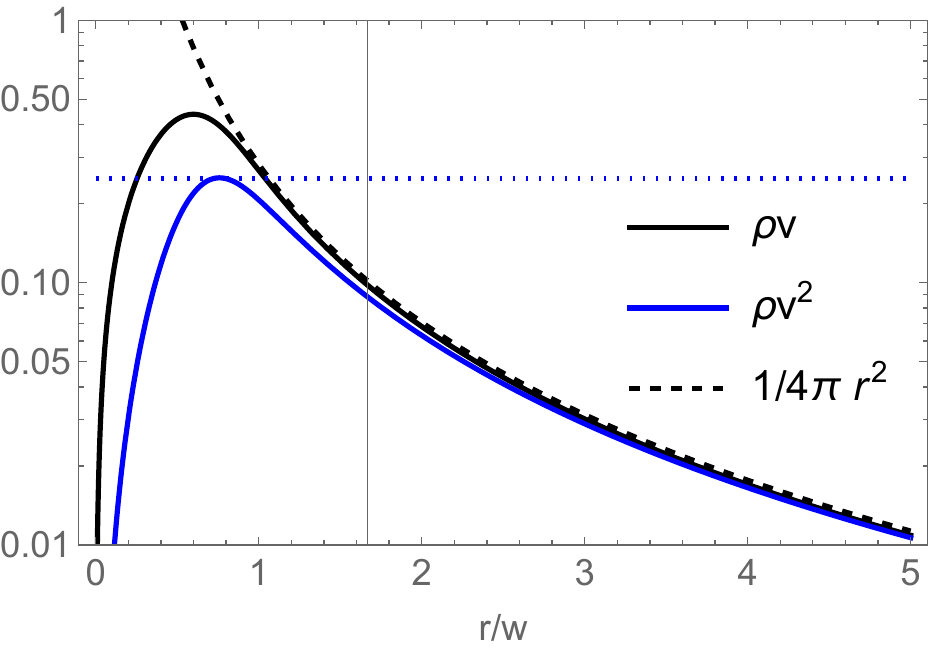}
\caption{Top: For the typical case of a gaussian mass deposition, the radial variation of mass (black) and momentum (blue) loss rate.
Bottom: Radial variation of mass (black) and momentum (blue) flux density.
If the local interstellar pressure were to exceed the peak (marked by horizontal dotted line) of the blue curve, 
the usual supersonic outflow solution would revert to a slower, subsonic ``breeze''.
}
\label{fig:fig7}
\end{center}
\end{figure}

\subsection{Incomplete wind thermalization}
\label{sec:sec52}

As discussed in the Introduction, a  fundamental question is the degree to which the winds within a cluster become thermalized by 
mutual collisions \citep{Lopez11,Rosen14,Lancaster21,Rosen22}.
The above analysis assumes they are {\em fully thermalized}, and so only have the {\em scalar} effects of adding mass and energy to the
intra-cluster medium.

More generally one would expect the momentum components of the individual winds 
along the outward radius vector of the cluster would impart an outward {\em vector}
momentum to the medium.

To explore this in the context of the present semi-analytic approach, let us  define a thermalization fraction $f$ for the mass source $s$.
The remaining  component $1-f$ then adds a direct momentum addition $(1-f) s V_w$ to the outflow momentum
eqn.\ (\ref{eq:momcons}), which now
becomes
\beq
\rho v \frac{dv}{dr} = - \frac{dP}{dr}  - f  \, s \, v \,  +\,  (1-f)s V_w
\, .
\label{eq:momcons2}
\eeq

The non-thermalized flow will retain the (lower) specific enthalpy of the wind $h_w$,
but if we ignore that effect and just note that the work from this momentum addition still adds 
a specific energy set by $V_w$, we can assume the Bernoulli relation (\ref{eq:bern})
still roughly holds.
Then again using eqns. (\ref{eq:rhov})-(\ref{eq:hdef}),  (\ref{eq:bern}), and (\ref{eq:momcons}),
and now assuming the monatomic case $\gamma=5/3$,
the modified form of eqn.\ (\ref{eq:eomgg}) becomes 
\beq
\frac{dv}{dr}= \frac{v}{r}\,  \frac{ 5 \sigma v (f v +(f-1) \vw )  + (\vwsq - v^2)(\sigma-2)}{V_w^2-4v^2}
\, . 
\label{eq:eom2}
\eeq
Taking as a  typical example the gaussian mass deposition model for $s(r)$ and $\sigma(r)$, the solution for speed $v(r)$
proceeds much as before.

To estimate an appropriate value for $f$, consider a local stellar wind source at a fixed position along
the sphere at intra-cluster radius $r$.
Any mass flowing inward from the source will eventually collide with other wind sources below, implying 
that at least half the total flow will be thermalized, $f > 1/2$.
For the outward wind outflow, its projection along the direction $r$ is just given by $\mu = \cos \theta$,
where $\theta$ is the wind co-latitude measured from its polar axis. 
Integration over the solid angle gives $\int_0^1 \mu d\mu = 1/2$, indicating that half of the wind momentum
over the upper hemisphere might impart a direct driving to the cluster outflow, without being thermalized.
Overall, this then suggests a net thermalization fraction $f=3/4$.

\begin{figure}
\begin{center}
\includegraphics[scale=0.5]{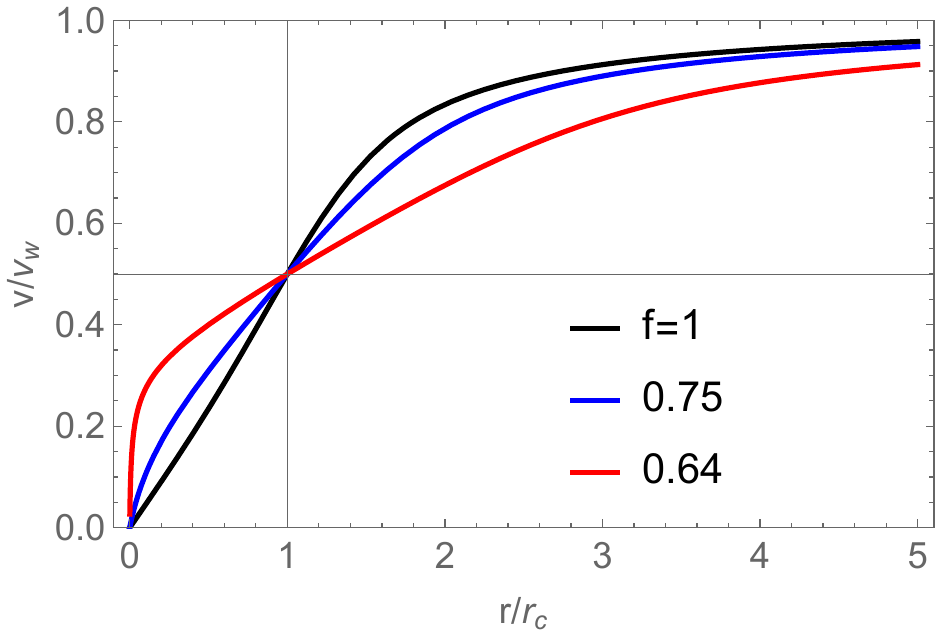}
\caption{For gaussian deposition models with the labeled values of the thermalization fraction $f$, comparison of speed vs. radius in critical units.
The preferred fraction $f=0.75$ (blue) has a slightly steeper initial acceleration than the full thermalization case $f=1$ (black).
The red curve shows a much steeper initial acceleration for the near-minimal thermalization $f=0.64$, with even smaller
values giving unbounded acceleration at the origin..
}
\label{fig:fig8}
\end{center}
\end{figure}

Figure \ref{fig:fig8} compares the resulting speed variation of the standard full thermalization case 
$f=1$ (black curve) with that for this reduced  thermalization models $f = 0.75$ (blue curve), along
with an ever lower case $f=0.64$ (red curve).  
The latter turns out to be near an allowed lower limit, with initially
steep acceleration followed by a more gradual radial increase.
This can be understood by noting that the lower thermalization reduces the drag effect while adding near the 
origin a relatively strong positive acceleration that scales with the stellar wind speed $V_w$.

The more preferred  model with $f=3/4$ (blue curve) also has a steeper initial acceleration, though it is much less dramatic,
with an overall speed $v(r)$ that is quite similar to the full thermalization case, $f=1$

Within this simplified semi-analytic picture\footnote{More general models might take the  thermalization fraction to have a radial decline from near unity at the cluster center. But accounting for this variable $f(r)$ in the energy eqn.\ (\ref{eq:eneq1}) would invalidate the Bernoulli form (\ref{eq:bern}), which is {\em essential} for our semi-analytic approach here.}, it thus seems that an incomplete thermalization of the source winds,  with instead  some direct momentum
addition to the cluster outflow,  might not dramatically alter the cluster velocity law $v(r)$ from what is obtained in full thermalization models.
But the analysis by \citet{Rosen14} indicates bubbles from cluster outflows expand more slowly than expected if all the winds are thermalized.
In the small thermalization limit, the mass-loaded solutions may tend to those developed by \citet{Smith96}, possibly with non-zero central speed.
Thus further studies with numerical simulations are needed to explore this issue of thermalization.

\subsection{Radiative losses in stellar wind termination shocks}
\label{sec:sec53}

Let us finally estimate the potential importance of radiative losses in the termination shock of individual WR winds
impacting into the ambient cluster medium.
As noted in eqn.\ (\ref{eq:rbdef}), the location of termination shock will depend on the ambient pressure of the intra-cluster medium.
Near the cluster center, we can use eqns.\ (\ref{eq:To}) and (\ref{eq:rhoo}) to show this pressure scales as
\beq
P_o = \rho k T_o/\mu \approx 0.024 \, N_{wr} \frac{{\dot M}_{wr}  V_w}{w^2}
\, ,
\label{eq:Po}
\eeq
where the ratio of the total cluster mass loss to that of  individual WR winds is set by the total number of WR stars, $N_{wr}$.
The associated wind momentum balance radius is thus set by 
\beq
r_o = \sqrt{\frac{{\dot M}_{wr} V_w}{4 \pi P_o}} \
\approx 1.8 \,  \frac{w}{\sqrt{N_{wr}}}
\, .
\label{eq:rodef}
\eeq
Assuming a standard factor 4 compression in the termination shock, the cooling efficiency is given by eqn.\ (\ref{eq:epsLam})
enhanced by a factor 16 to account for the density-squared dependence of radiative cooling,
\beq
\epsilon_{wr} =
 \frac{{16 \bar \Lambda}_o {\dot M} }{2 \pi r_o V_w^4} 
 \approx  1.9 \times 10^{-5} \sqrt{N_{wr}} \, \frac{{\dot M}_{-5}}{w_{pc} V_8^4} 
\, ,
\label{eq:epswr}
\eeq
where the latter scalings use a standard WR wind mass loss, ${\dot M}_{-5} \equiv {\dot M}_{wr}/10^{-5} M_\odot$/yr.
Away from the cluster center, the lower cluster pressure and thus larger termination distance should give an even smaller cooling ratio.

The upshot is that radiative cooling in wind termination shocks is likewise likely to lead to only small reduction in the thermal energy addition
from the stellar winds into the overall cluster.
As such, it does not seem to be a viable explanation for why the observed X-rays are found to be well below the predictions for 
these
thermalized 
cluster models \citep{Lopez11,Rosen14,Gupta18,Lancaster21}.

\section{Summary and future outlook}
\label{sec:sec6}

This paper has developed semi-analytic models for adiabatic outflows from stellar clusters
using more general treatments for the cumulative mass  deposition $s(r)$ 
from a  large population of mass-losing stars.
The models are  parameterized by 
the total wind mass-loss rate ${\dot M}$,
the energy-averaged wind flow speed $V_w$,
and
the mass deposition ratio $\sigma(r) \equiv r^3 s(r)/\int_0^r r'^2 s(r') dr'$.
Governing equations are derived for a general adiabatic index $\gamma$, 
but all solutions obtained here are for the case $\gamma=5/3$, applicable to a monatomic
ideal gas, leading here to the key critical point equation (\ref{eq:eomgg}).

Here is a summary list of key results:

\begin{itemize}

\item For any given mass deposition, our approach shows that for adiabatic expansion with the standard index $\gamma=5/3$,
 the critical/sonic radius can be obtained from the condition $\sigma(r_c)=3/4$.
Given the energy-averaged wind speed $V_w$, the associated critical/sonic speed for all mass deposition models is $v_c = c_{sc} = V_w/2$.
Integration inward/outward from the critical radius gives a much simpler and more stable velocity solution than the general `shooting method' used in the previous analysis by \citet{Silich11}.

\item We first apply this adiabatic flow analysis to previous models with mass loading that have a power-law \citep{Canto00,Rodriguez-Gonzalez07} or exponential \citep{Silich11} variation in radius, showing that the critical/sonic radius is set by the outer cutoff radius for the power-law models, but
given analytically at 4.07 times the e-fold size $w_e$ for the exponential model.
 The comparison of velocity solutions  in figure \ref{fig:fig1} shows that the power-law models have a diverging slope at {\color {red} the} sonic point, while the exponential 
 model has smooth, gradual transonic transition.
 All models have a terminal speed set by $V_w$,  again with a critical/sonic speed $V_w/2$.  

\item Inclusion of radiative cooling effects can be cast in terms of a `cooling speed' $V_\Lambda$, defined by the integral in eqn.\ (\ref{eq:VLamdef}).
The coupling of this with the resulting critical equation (\ref{eq:dvdrLam}) generally requires solution by a shooting method, as done by \citet{Silich11}.
But the scaling analysis in eqn. (\ref{eq:epsLam}) shows that such cooling is generally only a small, perturbative effect, characterized by a cooling efficiency $\epsilon_\Lambda \ll 1$.  
Even the most extreme model C of \citet{Silich11} has $\epsilon_\Lambda \approx 0.20$, leading to only a 10\% reduction in terminal speed.

\item Section \ref{sec:sec53} likewise shows that radiative cooling losses in the individual wind termination shocks should generally lead to only a small reduction in the wind energy deposition into the intra-cluster medium.

\item Neglecting radiative cooling, we thus derive (section \ref{sec:sec3}) purely {\em adiabatic} flow solutions for new models with zero gradient in mass deposition at the origin, including for a gaussian and for empirical power-law models derived by \citet{Plummer11} and \citet{Elson87}.
We show (figures \ref{fig:fig3} and \ref{fig:fig4}) that, when cast in terms of radius scaled by critical radius $r/r_c$, all deposition models have similar 
variations in velocity,
constrained to increase from zero speed at the origin, through $V_w/2$ at the critical radius, and asymptoting to $V_w$ at large radii. (However the \citet{Elson87} model with index near the limiting value $\gamma_e \rightarrow 2$ has very slow, logarithmic increase toward this large-radius asymptote.)

\item For all models, the central temperature  $T_o$ is  set by the wind speed $V_w$ through eqn.\ (\ref{eq:Todef}),  while the central density $\rho_o$ scales also with mass loss rate ${\dot M}$ and cluster half-width $w$, modified by the ratio $r_c/w$ for the various deposition models (eqn.\ (\ref{eq:rhoo})).
The ratio of associated base cooling luminosity to wind kinetic energy luminosity is effectively set by the cooling efficiency,  $L_o/L_{KE} \sim \epsilon_\Lambda$ (cf.\ eqns. (\ref{eq:epsLam}) and (\ref{eq:LobLke})).

\item This radiative cooling scaling also allows derivation of scaling forms for X-ray luminosity and surface brightness,
parameterized here by the X-ray energy in temperature units $T_x$ (see figure \ref{fig:fig6}).
When surface brightness is normalized to the central value, $I_x (p,T_x)/I_x (p=0,T_x)$, and corrected by a factor 
$e^{T_x/T_o}$, the variations with offset $p$ are similar for all deposition models and all $T_x$.
Likewise, scaling the integrated X-ray luminosities scaled by this factor $ e^{T_x/T_o}$ results in values that are a factor 0.4-0.7 the base radiative luminosity $L_o$.

\item Comparison of the outflow momentum flux vs. the interstellar medium pressure $P_{\rm ISM}$ indicates supersonic outflow into
a strong termination shock should be the general outcome, with inhibition into subsonic ``breeze'' outflows only possible
for ISM pressures much higher than typical for our galaxy.

\item An initial semi-analyic account of the effect of incomplete wind thermalization indicates that this may have only a minor  effect on the wind outflow, but further analysis that accounts for a sparse distribution of wind sources is still needed.

\end{itemize}

A key issue for future work regards this last point on the sparseness of the wind sources, which leads to reduced  thermalization
 and instead a more directed cumulative outflow from the individual stellar winds.
For example, the recent simulations by \citet{Vieu24} of the cluster Cygnus OB2 find that the wind mass sources are so sparse that
there is not even a clearly formed cluster termination shock into the local ISM.
In  MHD simulations of interacting winds from young massive stellar clusters, \citet{Badmaev22} find filamentary magnetic field structures that could be important for cosmic ray acceleration and high-energy gamma-ray emission.

To clarify this transition from the idealized fully thermalized picture assumed here, it would be interesting to carry out stylized numerical 
models with a fixed array of stellar wind sources varying from closely packed to relatively sparse \citep{Scherer18}.

For analysis of X-ray emission, future studies should examine the potential role of non-equilibrium ionization and incomplete equilibration of
electron and proton temperatures in the relatively low-density of the intra-cluster medium \citep{Raga01, Ji06,Rosen14, Muno04,Kavanagh20}.
Metallicity effects were studied by \citet{Silich04}. Turbulent mixing and associated instabilities can also have important effects on X-ray emission \citep{Rosen14,Gallegos-Garcia20, Lancaster21}.
Another issue regards the effect of reduced hydrogen abundances given that a principal wind source is likely to be from classical WR stars with  stripped Hydrogen envelopes, though the strongest winds can be from the very massive WNH stars that still have hydrogen.

Finally, to model stellar clusters near galactic centers, it would be interesting to include the
gravitational effects of the central super-massive black hole,  accounting also for a central sink from
mass accretion \citep{Silich08} and associated feedback of the resulting emission of gas and radiation.

But overall, the generalized semi-analytic outflow solutions derived here should provide a useful basis for interpreting results from more 
detailed numerical hydrodynamics simulation of actual stellar clusters.

\section*{Acknowledgements}

I thank the anonymous referee for many constructive comments and suggestions that helped substantially improve this paper from the originally submitted version. I also thank the organizers of the recent TOSCA 2024 conference, where much of the impetus for this paper originated.
I particularly thank Anna Rosen for pointing out the \citet{Canto00} paper. I also thank Jonathan Mackey, Julian Pittard,
Chris Russell, Andreas Sander,  Jacco Vink, and Jorick Vink for helpful comments on various drafts.
I acknowledge travel and general support from the Bartol Research Institute, and from  NASA ATP grant number 80NSSC22K0628.

\section*{Data Availability Statement}

 No new data were generated or analysed in support of this research.

\bibliographystyle{mn2e}
\bibliography{OwockiS}

\begin{thebibliography}{}

\bibitem[\protect\citeauthoryear{{Aharonian}, {Yang} \& {de O{\~n}a
  Wilhelmi}}{{Aharonian} et~al.}{2019}]{Aharonian19}
{Aharonian} F.,  {Yang} R.,    {de O{\~n}a Wilhelmi} E.,  2019, Nature
  Astronomy, 3, 561

\bibitem[\protect\citeauthoryear{{Avedisova}}{{Avedisova}}{1972}]{Avedisova72}
{Avedisova} V.~S.,  1972, \sovast, 15, 708

\bibitem[\protect\citeauthoryear{{Badmaev}, {Bykov} \& {Kalyashova}}{{Badmaev}
  et~al.}{2022}]{Badmaev22}
{Badmaev} D.~V.,  {Bykov} A.~M.,    {Kalyashova} M.~E.,  2022, \mnras, 517,
  2818

\bibitem[\protect\citeauthoryear{{Cant{\'o}}, {Raga} \&
  {Rodr{\'\i}guez}}{{Cant{\'o}} et~al.}{2000}]{Canto00}
{Cant{\'o}} J.,  {Raga} A.~C.,    {Rodr{\'\i}guez} L.~F.,  2000, \apj, 536, 896

\bibitem[\protect\citeauthoryear{{Chevalier} \& {Clegg}}{{Chevalier} \&
  {Clegg}}{1985}]{Chevalier85}
{Chevalier} R.~A.,  {Clegg} A.~W.,  1985, \nat, 317, 44

\bibitem[\protect\citeauthoryear{{Dyson} \& {de Vries}}{{Dyson} \& {de
  Vries}}{1972}]{Dyson72}
{Dyson} J.~E.,  {de Vries} J.,  1972, \aap, 20, 223

\bibitem[\protect\citeauthoryear{{Elson}, {Fall} \& {Freeman}}{{Elson}
  et~al.}{1987}]{Elson87}
{Elson} R. A.~W.,  {Fall} S.~M.,    {Freeman} K.~C.,  1987, \apj, 323, 54

\bibitem[\protect\citeauthoryear{{Gallegos-Garcia}, {Burkhart}, {Rosen},
  {Naiman} \& {Ramirez-Ruiz}}{{Gallegos-Garcia}
  et~al.}{2020}]{Gallegos-Garcia20}
{Gallegos-Garcia} M.,  {Burkhart} B.,  {Rosen} A.~L.,  {Naiman} J.~P.,
  {Ramirez-Ruiz} E.,  2020, \apjl, 899, L30

\bibitem[\protect\citeauthoryear{{Gupta}, {Nath}, {Sharma} \&
  {Eichler}}{{Gupta} et~al.}{2018}]{Gupta18}
{Gupta} S.,  {Nath} B.~B.,  {Sharma} P.,    {Eichler} D.,  2018, \mnras, 473,
  1537

\bibitem[\protect\citeauthoryear{{Harper-Clark} \& {Murray}}{{Harper-Clark} \&
  {Murray}}{2009}]{Harper-Clark09}
{Harper-Clark} E.,  {Murray} N.,  2009, \apj, 693, 1696

\bibitem[\protect\citeauthoryear{{Ji}, {Wang} \& {Kwan}}{{Ji}
  et~al.}{2006}]{Ji06}
{Ji} L.,  {Wang} Q.~D.,    {Kwan} J.,  2006, \mnras, 372, 497

\bibitem[\protect\citeauthoryear{{Kavanagh}}{{Kavanagh}}{2020}]{Kavanagh20}
{Kavanagh} P.~J.,  2020, \apss, 365, 6

\bibitem[\protect\citeauthoryear{{Koo} \& {McKee}}{{Koo} \&
  {McKee}}{1992}]{Koo92}
{Koo} B.-C.,  {McKee} C.~F.,  1992, \apj, 388, 93

\bibitem[\protect\citeauthoryear{{Lancaster}, {Ostriker}, {Kim} \&
  {Kim}}{{Lancaster} et~al.}{2021}]{Lancaster21}
{Lancaster} L.,  {Ostriker} E.~C.,  {Kim} J.-G.,    {Kim} C.-G.,  2021, \apj,
  914, 89

\bibitem[\protect\citeauthoryear{{Lopez}, {Krumholz}, {Bolatto}, {Prochaska} \&
  {Ramirez-Ruiz}}{{Lopez} et~al.}{2011}]{Lopez11}
{Lopez} L.~A.,  {Krumholz} M.~R.,  {Bolatto} A.~D.,  {Prochaska} J.~X.,
  {Ramirez-Ruiz} E.,  2011, \apj, 731, 91

\bibitem[\protect\citeauthoryear{{McKee} \& {Ostriker}}{{McKee} \&
  {Ostriker}}{1977}]{McKee77}
{McKee} C.~F.,  {Ostriker} J.~P.,  1977, \apj, 218, 148

\bibitem[\protect\citeauthoryear{{Morlino}, {Blasi}, {Peretti} \&
  {Cristofari}}{{Morlino} et~al.}{2021}]{Morlino21}
{Morlino} G.,  {Blasi} P.,  {Peretti} E.,    {Cristofari} P.,  2021, \mnras,
  504, 6096

\bibitem[\protect\citeauthoryear{{Muno}, {Arabadjis}, {Baganoff}, {Bautz},
  {Brandt}, {Broos}, {Feigelson}, {Garmire}, {Morris} \& {Ricker}}{{Muno}
  et~al.}{2004}]{Muno04}
{Muno} M.~P.,  {Arabadjis} J.~S.,  {Baganoff} F.~K.,  {Bautz} M.~W.,  {Brandt}
  W.~N.,  {Broos} P.~S.,  {Feigelson} E.~D.,  {Garmire} G.~P.,  {Morris} M.~R.,
     {Ricker} G.~R.,  2004, \apj, 613, 1179

\bibitem[\protect\citeauthoryear{{Pandey}, {Lopez}, {Rosen}, {Thompson},
  {Linden} \& {Blackstone}}{{Pandey} et~al.}{2024}]{Pandey24}
{Pandey} P.,  {Lopez} L.~A.,  {Rosen} A.~L.,  {Thompson} T.~A.,  {Linden} T.,
   {Blackstone} I.,  2024, \apj, 976, 98

\bibitem[\protect\citeauthoryear{{Pascale} \& {Dai}}{{Pascale} \&
  {Dai}}{2024}]{Pascale24}
{Pascale} M.,  {Dai} L.,  2024, \apj, 976, 166

\bibitem[\protect\citeauthoryear{{Peron}, {Morlino}, {Gabici}, {Amato},
  {Purushothaman} \& {Brusa}}{{Peron} et~al.}{2024}]{Peron24}
{Peron} G.,  {Morlino} G.,  {Gabici} S.,  {Amato} E.,  {Purushothaman} A.,
  {Brusa} M.,  2024, \apjl, 972, L22

\bibitem[\protect\citeauthoryear{{Pikel'Ner}}{{Pikel'Ner}}{1968}]{Pikelner68}
{Pikel'Ner} S.~B.,  1968, \aplett, 2, 97

\bibitem[\protect\citeauthoryear{{Plummer}}{{Plummer}}{1911}]{Plummer11}
{Plummer} H.~C.,  1911, \mnras, 71, 460

\bibitem[\protect\citeauthoryear{{Puls}, {Vink} \& {Najarro}}{{Puls}
  et~al.}{2008}]{Puls08}
{Puls} J.,  {Vink} J.~S.,    {Najarro} F.,  2008, \aapr, 16, 209

\bibitem[\protect\citeauthoryear{{Raga}, {Vel{\'a}zquez}, {Cant{\'o}},
  {Masciadri} \& {Rodr{\'\i}guez}}{{Raga} et~al.}{2001}]{Raga01}
{Raga} A.~C.,  {Vel{\'a}zquez} P.~F.,  {Cant{\'o}} J.,  {Masciadri} E.,
  {Rodr{\'\i}guez} L.~F.,  2001, \apjl, 559, L33

\bibitem[\protect\citeauthoryear{{Rodr{\'\i}guez-Gonz{\'a}lez}, {Cant{\'o}},
  {Esquivel}, {Raga} \& {Vel{\'a}zquez}}{{Rodr{\'\i}guez-Gonz{\'a}lez}
  et~al.}{2007}]{Rodriguez-Gonzalez07}
{Rodr{\'\i}guez-Gonz{\'a}lez} A.,  {Cant{\'o}} J.,  {Esquivel} A.,  {Raga}
  A.~C.,    {Vel{\'a}zquez} P.~F.,  2007, \mnras, 380, 1198

\bibitem[\protect\citeauthoryear{{Rogers} \& {Pittard}}{{Rogers} \&
  {Pittard}}{2013}]{Rogers13}
{Rogers} H.,  {Pittard} J.~M.,  2013, \mnras, 431, 1337

\bibitem[\protect\citeauthoryear{{Rosen}}{{Rosen}}{2022}]{Rosen22}
{Rosen} A.~L.,  2022, \apj, 941, 202

\bibitem[\protect\citeauthoryear{{Rosen}, {Lopez}, {Krumholz} \&
  {Ramirez-Ruiz}}{{Rosen} et~al.}{2014}]{Rosen14}
{Rosen} A.~L.,  {Lopez} L.~A.,  {Krumholz} M.~R.,    {Ramirez-Ruiz} E.,  2014,
  \mnras, 442, 2701

\bibitem[\protect\citeauthoryear{{Scherer}, {Noack}, {Kleimann}, {Fichtner} \&
  {Weis}}{{Scherer} et~al.}{2018}]{Scherer18}
{Scherer} K.,  {Noack} A.,  {Kleimann} J.,  {Fichtner} H.,    {Weis} K.,  2018,
  \aap, 616, A115

\bibitem[\protect\citeauthoryear{{Schure}, {Kosenko}, {Kaastra}, {Keppens} \&
  {Vink}}{{Schure} et~al.}{2009}]{Schure09}
{Schure} K.~M.,  {Kosenko} D.,  {Kaastra} J.~S.,  {Keppens} R.,    {Vink} J.,
  2009, \aap, 508, 751

\bibitem[\protect\citeauthoryear{{Silich}, {Bisnovatyi-Kogan}, {Tenorio-Tagle}
  \& {Mart{\'\i}nez-Gonz{\'a}lez}}{{Silich} et~al.}{2011}]{Silich11}
{Silich} S.,  {Bisnovatyi-Kogan} G.,  {Tenorio-Tagle} G.,
  {Mart{\'\i}nez-Gonz{\'a}lez} S.,  2011, \apj, 743, 120

\bibitem[\protect\citeauthoryear{{Silich}, {Tenorio-Tagle} \&
  {A{\~n}orve-Zeferino}}{{Silich} et~al.}{2005}]{Silich05}
{Silich} S.,  {Tenorio-Tagle} G.,    {A{\~n}orve-Zeferino} G.~A.,  2005, \apj,
  635, 1116

\bibitem[\protect\citeauthoryear{{Silich}, {Tenorio-Tagle} \&
  {Hueyotl-Zahuantitla}}{{Silich} et~al.}{2008}]{Silich08}
{Silich} S.,  {Tenorio-Tagle} G.,    {Hueyotl-Zahuantitla} F.,  2008, \apj,
  686, 172

\bibitem[\protect\citeauthoryear{{Silich}, {Tenorio-Tagle} \&
  {Rodr{\'\i}guez-Gonz{\'a}lez}}{{Silich} et~al.}{2004}]{Silich04}
{Silich} S.,  {Tenorio-Tagle} G.,    {Rodr{\'\i}guez-Gonz{\'a}lez} A.,  2004,
  \apj, 610, 226

\bibitem[\protect\citeauthoryear{{Smith}}{{Smith}}{1996}]{Smith96}
{Smith} S.~J.,  1996, \apj, 473, 773

\bibitem[\protect\citeauthoryear{{Stevens}, {Blondin} \& {Pollock}}{{Stevens}
  et~al.}{1992}]{Stevens92}
{Stevens} I.~R.,  {Blondin} J.~M.,    {Pollock} A.~M.~T.,  1992, \apj, 386, 265

\bibitem[\protect\citeauthoryear{{ud-Doula}, {Owocki}, {Townsend}, {Petit} \&
  {Cohen}}{{ud-Doula} et~al.}{2014}]{Uddoula14}
{ud-Doula} A.,  {Owocki} S.,  {Townsend} R.,  {Petit} V.,    {Cohen} D.,  2014,
  \mnras, 441, 3600

\bibitem[\protect\citeauthoryear{{Vieu}, {Larkin}, {H{\"a}rer}, {Reville},
  {Sander} \& {Ramachandran}}{{Vieu} et~al.}{2024}]{Vieu24}
{Vieu} T.,  {Larkin} C.~J.~K.,  {H{\"a}rer} L.,  {Reville} B.,  {Sander}
  A.~A.~C.,    {Ramachandran} V.,  2024, \mnras, 532, 2174

\bibitem[\protect\citeauthoryear{{Weaver}, {McCray}, {Castor}, {Shapiro} \&
  {Moore}}{{Weaver} et~al.}{1977}]{Weaver77}
{Weaver} R.,  {McCray} R.,  {Castor} J.,  {Shapiro} P.,    {Moore} R.,  1977,
  \apj, 218, 377

\bibitem[\protect\citeauthoryear{{Westmoquette}, {Bastian}, {Smith}, {Seth},
  {Gallagher} J.~S., {O'Connell}, {Ryon}, {Silich}, {Mayya},
  {Mu{\~n}oz-Tu{\~n}{\'o}n} \& {Rosa Gonz{\'a}lez}}{{Westmoquette}
  et~al.}{2014}]{Westmoquette14}
{Westmoquette} M.~S.,  {Bastian} N.,  {Smith} L.~J.,  {Seth} A.~C.,
  {Gallagher} J.~S. I.,  {O'Connell} R.~W.,  {Ryon} J.~E.,  {Silich} S.,
  {Mayya} Y.~D.,  {Mu{\~n}oz-Tu{\~n}{\'o}n} C.,    {Rosa Gonz{\'a}lez} D.,
  2014, \apj, 789, 94

\end{thebibliography}

\end{document}